\documentclass[acmsmall]{acmart}
\usepackage{graphicx}
\usepackage{subcaption}
\usepackage{multirow}

\AtBeginDocument{%
  \providecommand\BibTeX{{%
    \normalfont B\kern-0.5em{\scshape i\kern-0.25em b}\kern-0.8em\TeX}}}

\setcopyright{acmlicensed}
\copyrightyear{2024}
\acmYear{2024}

\acmConference[CHI '24]{
Designing (with) AI for Wellbeing}{May,
  2024}{Honolulu, HI}
\acmISBN{978-1-4503-XXXX-X/24/05}

\newcommand{\sys}{PITCH}

\newcommand{\out}[1]{{#1}}

\newcommand{\summary}[1]{\out{{\textcolor{blue}{\textbf{ }}}}}

\begin{document}

\title{\sys{}: Productivity and Mental Well-being Coaching through Daily Conversational Interaction}


\author{Adnan Abbas}
\affiliation{%
  \institution{Virginia Tech}
  \city{Blacksburg}
  \country{USA}}
\email{adnana99@vt.edu}

\author{Sang Won Lee}
\affiliation{%
  \institution{Virginia Tech}
  \city{Blacksburg}
  \country{USA}
}

\renewcommand{\shortauthors}{Abbas and Lee.}

\begin{abstract}
Efficient task planning is essential for productivity and mental well-being, yet individuals often struggle to create realistic plans and reflect upon their productivity. 
Leveraging the advancement in artificial intelligence (AI), conversational agents have emerged as a promising tool for enhancing productivity. 
Our work focuses on externalizing plans through conversation, aiming to solidify intentions and foster focused action, thereby positively impacting their productivity and mental well-being. 
We share our plan of designing a conversational agent to offer insightful questions and reflective prompts for increasing plan adherence by leveraging the social interactivity of natural conversations. 
Previous studies have shown the effectiveness of such agents, but many interventions remain static, leading to decreased user engagement over time. 
To address this limitation, we propose a novel rotation and context-aware prompting strategy, providing users with varied interventions daily. Our system, \textit{\sys{}}, utilizes large language models (LLMs) to facilitate externalization and reflection on daily plans. 
Through this study, we investigate the impact of externalizing tasks with conversational agents on productivity and mental well-being, and the effectiveness of a rotation strategy in maintaining user engagement.

\end{abstract}



\maketitle

\section{Introduction}
Planning and keeping track of tasks is crucial for productivity and users’ mental well-being. However, we often underestimate the time it takes to complete tasks~\cite{buehler1994exploring} or fail to plan at all, leading to low productivity and stress~\cite{nihZeigarniksSleepless, AdamsFamilyStress, KimaniAMBER}. There are several task management tools for planning and keeping track of tasks, for e.g., calendars, to-do lists, project management software, notebooks, email, scraps of paper, and word documents~\cite{AhmetogluToPlanOrNotToPlan, HaratyHow, BernsteinInfoScraps}. However, despite such strategies, users still struggle with getting into the habit of planning~\cite{HaratyHow} and making realistic plans~\cite{BUEHLER20101}. Time management literature suggests that short-term planning behaviors show the most significant relationship to one's desired outcomes~\cite{TimeManagementReview, KearnsIsItTime, LundLessIsMore}. Our work focuses on how to facilitate workers' daily planning, or time management planning (TMP), which is defined as follows: determining tasks to be performed on a particular day, prioritizing tasks, and associating an approximate completion time with each task~\cite{ParkeTMP}. With the advancement of artificial intelligence and large-scale models, researchers have studied the potential of creating an intelligent agent that a worker can converse with regards to TMP~\cite{KimaniAMBER, GroverTwoAgents, KocielnikROBOTA}. 

Externalizing plans with a conversational agent could help solidify intentions and make them more deliberate. Simple reminders about one's plans, even before it is time to perform them, have the power to increase the likelihood of following the externalized plan~\cite{WicaksonoPlansDeliberate}. 
Using conversation to support externalization can be a promising approach due to the social interactivity that can emerge in a natural conversation. 
For example, personal coaches often employ a technique of repeated inquiries to uncover concealed motivations~\cite{LeePersonalCoach}. 
This method suggests that such conversations can prompt contemplative and meta-cognitive thinking~\cite{KocielnikReflectionCompanion}. 
Previous works experimented with different Conversational Agents (CAs) prototypes to support daily planning through conversation~\cite{KimaniAMBER, GroverTwoAgents}, to facilitate reflection and goal-setting~\cite{KocielnikReflectionCompanion, MeyerSOFTWAREDEVS, KocielnikROBOTA}, and supporting people to gain awareness and reflect on their health and activity data~\cite{Kinnafick, LeeVincent}. Commercial apps such as Rosebud~\cite{rosebudRosebudJournal} and Lark~\cite{larkHomeLark} serve as a “Chatbot,” actively initiating conversations with people by asking about their daily activities and well-being. 
Some of the findings from these studies suggest the success of employing a conversational agent in TMP; users feel more productive and satisfied with an emotionally expressive virtual agent~\cite{GroverTwoAgents} and follow their exercise plans consistently~\cite{LuoTandemTrack}.

While these works investigated people’s TMP behaviors with the support of a chatbot, most of them employ a static intervention in terms of conversational interaction, i.e., prompting the same set of questions repeatedly. 
Static interventions aimed at behavioral regulation are shown to decline in effectiveness over time as users begin to ignore them~\cite{KovacsStaticIntervention}. For instance, in the FitTrack study~\cite{FitTrackLoseInterest}, some subjects mentioned their lack of motivation to use the system because of its repetitive dialog.
One exception to the static intervention is work by Kocielnik et al., where different conversational mini-dialogues and follow-ups are used to initiate reflection on physical activity data~\cite{KocielnikReflectionCompanion}. The mini-dialogues are generated from a set of 275 prompts gathered through user workshops. However, these mini-dialogues are used only in the context of data sense-making in the fitness domain.

We aim to introduce a \textit{rotation} and \textit{context-aware prompting} strategy, where users experience different interventions every day through different prompts/questions with the use of Large Language Models (LLMs).
The advancement in natural language processing (NLP), especially the recent surge of LLMs, has opened up exciting opportunities for designers and developers to customize chatbots that engage people in more natural and fluent conversations~\cite{seo2024chacha, HanInterviewChatbots, HarringtonOldPeopleLLM, wei2023leveraging}.  
Rotation strategy is defined as cycling through different approaches over time to maintain sustained effectiveness and engagement, just like a human coach iterates with novel strategies.
Context-aware prompting means the system personalizes questions according to the user's context, which involves users' previous responses.  
Novel and relatable questions coming from rotated and context-aware prompting strategies can increase effectiveness since novelty can influence encoding online information into long-term memory, which, in turn, may raise awareness of behavioral changes~\cite{KormiNovelty}. Moreover, systems that personalize interventions~\cite{KapteinPersonalize} have found positive effects in behavioral change.

Currently, we are in the process of developing an intelligent LLM-based conversational system (\sys{}, Personal and Intelligent Task-management Conversation Helper) that can support users in externalization and reflection about their daily plans using a rotational strategy. We aim to answer the following research questions for our study:

\begin{itemize}
    \item \textbf{RQ1:} How does externalization of tasks and follow-up by/with an external agent help users with productivity and mental well-being?
\end{itemize}
\begin{itemize}
    \item \textbf{RQ2:} How does the rotation strategy of prompting different questions every day help maintain users' interests in the conversational system for reflection?
\end{itemize}

\section{\sys{} - Conversational Agent for Reflecting on Task Management}

We conducted a pilot study to understand the effectiveness of using a conversational system in supporting users with externalization and reflection on their daily tasks. Based on the findings from the pilot study and literature review, we came up with the following design goals for \sys{} to answer the research questions:

\subsection{Design Goals}
\subsubsection{Help users to externalize their plans through conversation at the beginning of their day} 
As discussed in the Introduction, users struggle with planning their day and often miscalculate their plans, resulting in stress and lower productivity~\cite{AhmetogluToPlanOrNotToPlan}. 
We aim to bridge the gap between users' work expectations and reality by making their plans well-intentioned and achievable. Therefore, we need to design a chat interface for the exchange between the user and the agent. 
We will prompt workers with a task-management question twice per day: morning and evening. 
The morning question will be generated with a randomly selected goal that is relevant to either productivity or mental-well being and will be focused on short-term planning.
Based on the user's response, the system will ask further questions to help workers externalize and incorporate the task into their day's routine.
We also want to tailor the timing of the morning check-in according to the user's schedule so it is easier for users to adopt the habit. An example of morning interaction at the start of the user's day can be seen in Figure~\ref{fig: Scenario 1 - morning} and \ref{fig: Scenario 2 - morning}.

\subsubsection{Facilitate personalized reflection on users’ externalized plan and detachment from work through context-aware prompting}

\begin{figure*}[t!]
    \centering
    \begin{subfigure}[t]{0.5\textwidth}
        \centering
        \includegraphics[width=1.0\textwidth]{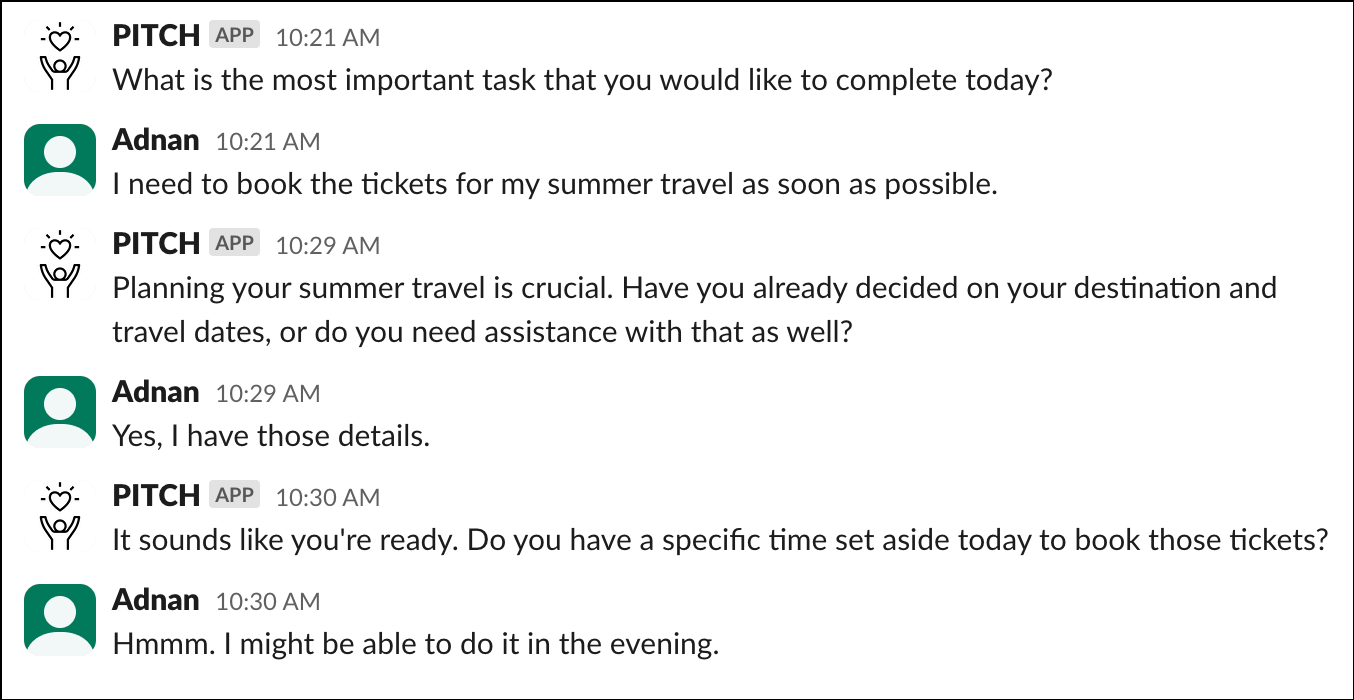}
        \caption{ Morning conversation}
        \label{fig: Scenario 1 - morning}
    \end{subfigure}%
    ~ 
    \begin{subfigure}[t]{0.5\textwidth}
        \centering
        \includegraphics[width=1.0\textwidth]{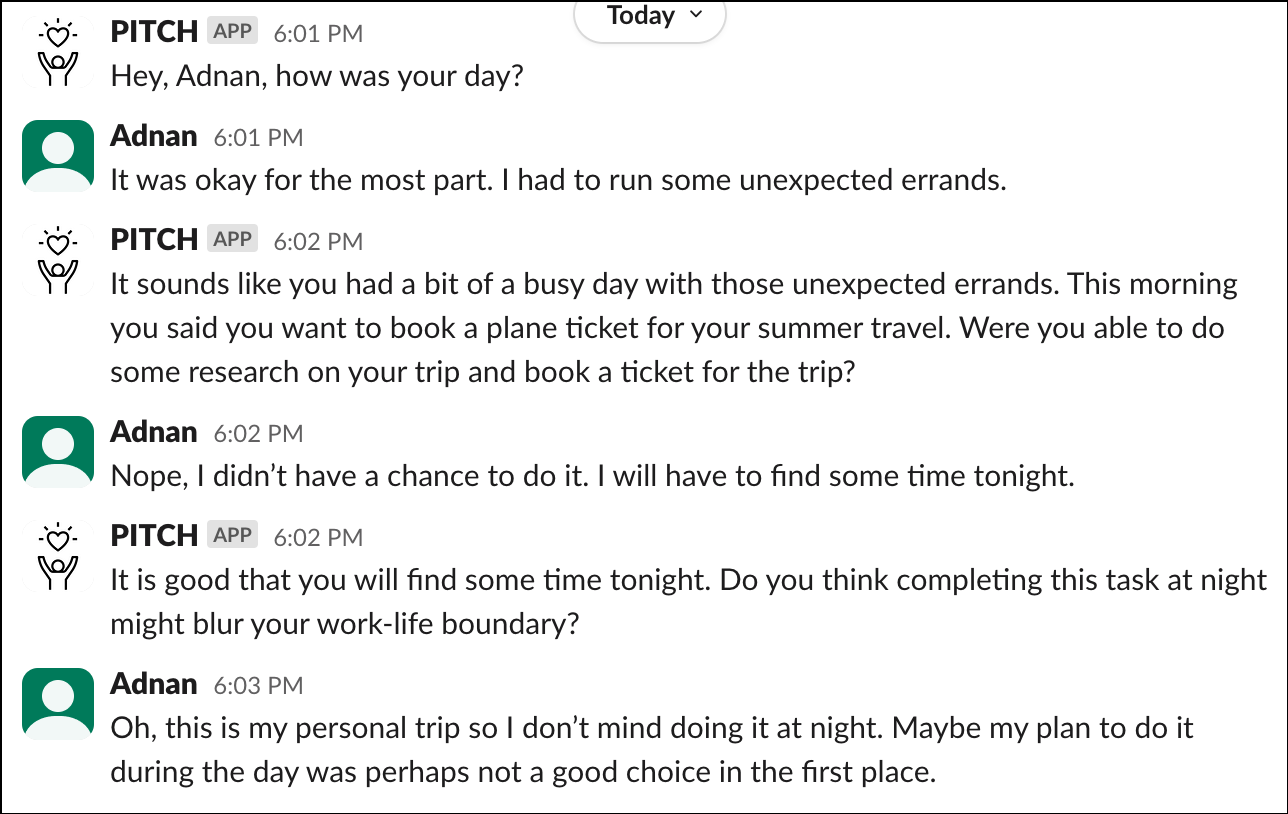}
        \caption{ Evening conversation}
        \label{fig: Scenario 1 - evening}
    \end{subfigure}
    \caption{Scenario 1 - Morning and Evening example conversation}
    \label{fig:Scenario 1}
\end{figure*}

\begin{figure*}[t!]
    \centering
    \begin{subfigure}[t]{0.5\textwidth}
        \centering
        \includegraphics[width=1.0\textwidth]{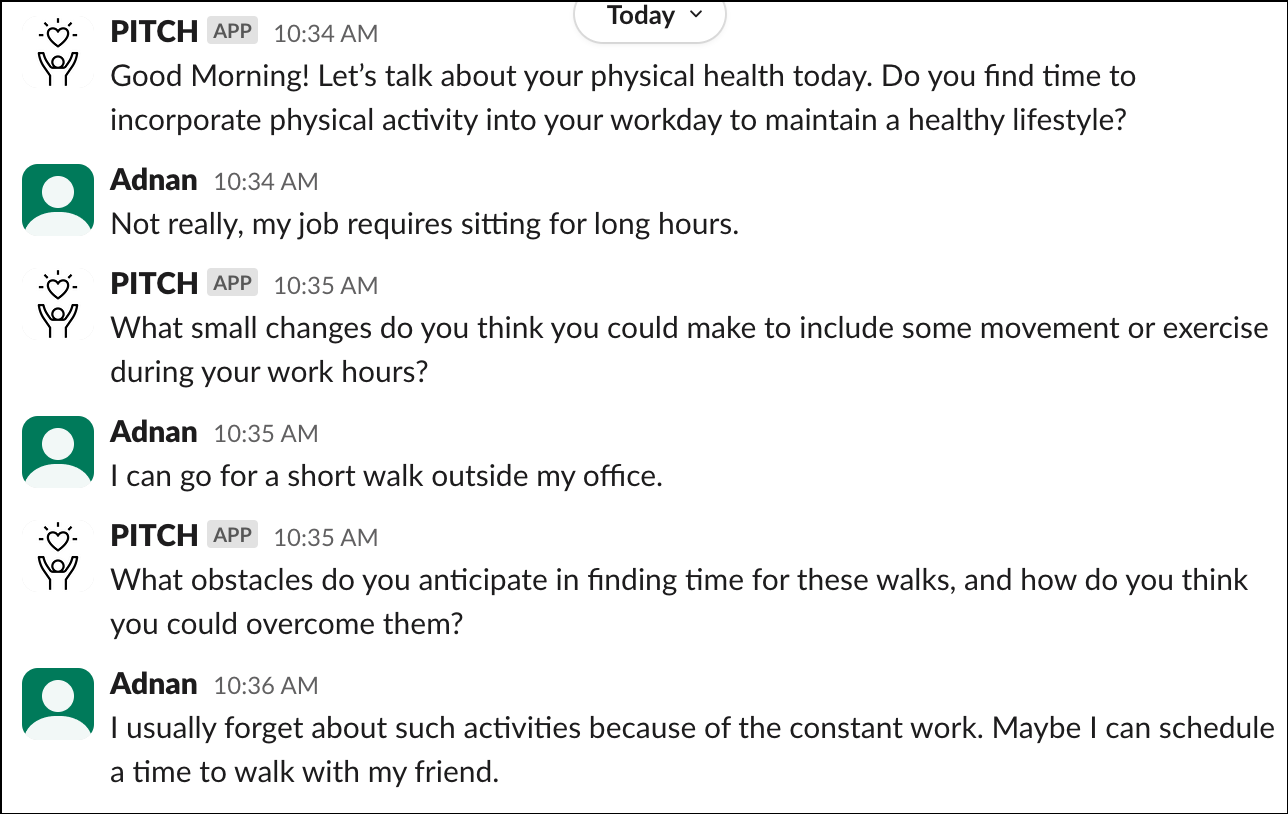}
        \caption{ Morning conversation}
        \label{fig: Scenario 2 - morning}
    \end{subfigure}%
    ~ 
    \begin{subfigure}[t]{0.5\textwidth}
        \centering
        \includegraphics[width=1.0\textwidth]{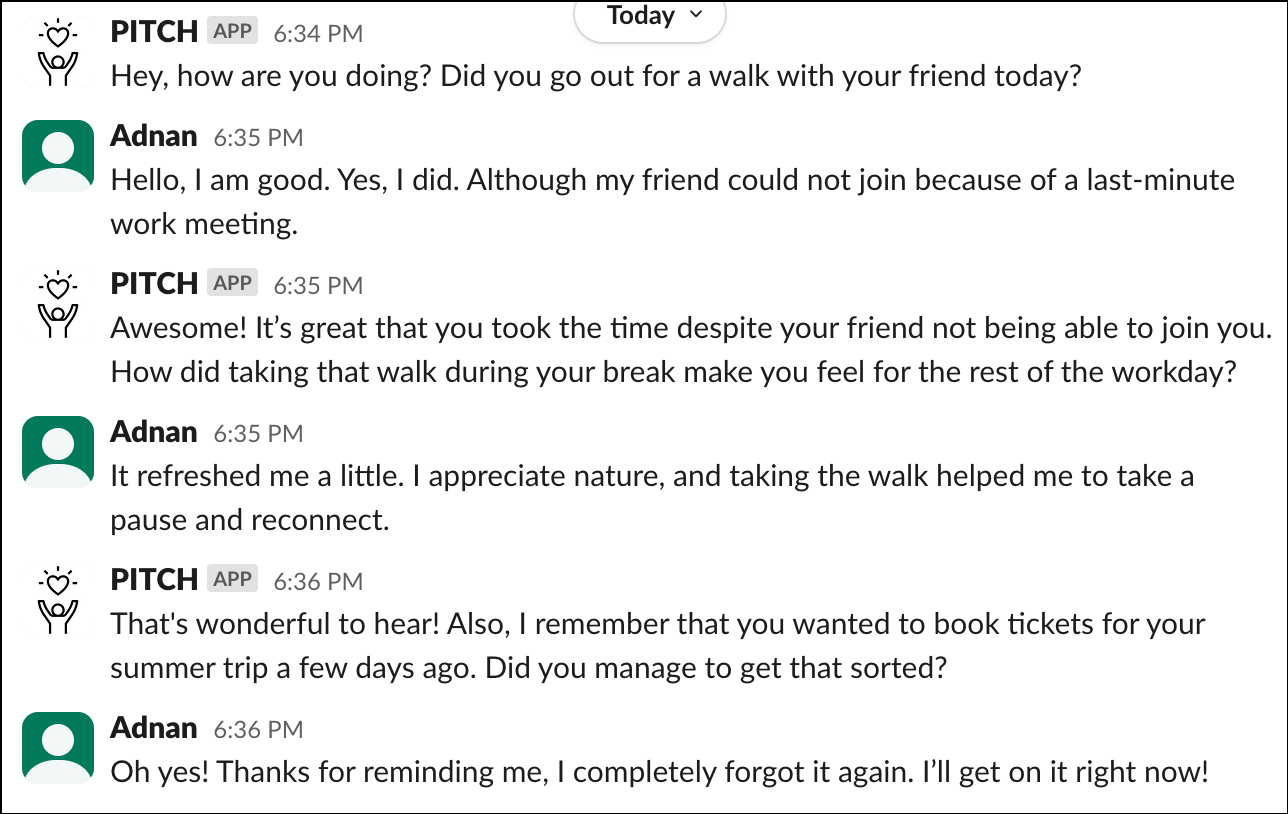}
        \caption{ Evening conversation}
        \label{fig: Scenario 2 - evening}
    \end{subfigure}
    \caption{Scenario 2 - Morning and Evening example conversation}
    \label{fig:Scenario 2}
\end{figure*}

Reflection of time spent can reduce absolute time on unimportant activities as indicated by the user studies~\cite{MongeRoffarello, focusboosterappFocusBooster}. Recent research suggests that the psychological benefits of disclosure and reflection with an agent are similar to reflection with another human~\cite{HoReflectionWithAgent}. Moreover, reflecting on one's plans at the end of one's day can help psychologically detach from work and support a healthy work-life balance~\cite{WilliamsDetachment, KimaniAMBER, SmitDetachment}. However, users need a nudge to reflect upon their activities as reflection does not come naturally to humans~\cite{FleckReflectingOnReflection}. Encouraging reflection on deeper levels, it is useful to ask contextual questions from users and doing this exercise with an agent can allow one to see more than one could possibly see alone~\cite{FleckReflectingOnReflection}. 
Therefore, we need to implement a conversation storage and retrieval module that can ask personalized reflective questions to users at the end of their day. 
We will create prompts that take in the stored morning conversation's context to generate personalized reflection questions that can help a worker understand the time spent during the day. 
Similar to the morning check-in, we also want to tailor the timing of the evening reflection according to the user's schedule so it is easier for users to adopt the habit.  An example interaction at the end of the user's day can be seen in Figure~\ref{fig: Scenario 1 - evening} and \ref{fig: Scenario 2 - evening}.

\renewcommand{\arraystretch}{1.25}

\subsubsection{Maintain user's interest by introducing randomness in externalization inquiry and reflection prompts}

As mentioned in the Introduction, users can lose interest in behavioral change systems because of static interventions. Therefore, we aim to achieve \textit{rotational strategy}~\cite{KovacsStaticIntervention} by asking different prompts/questions everyday. Another way to incorporate unpredictability in our system is by leveraging temporal randomness on user's contextual information as a seed for prompt generation, for instance, the system asks the user “You wanted to book tickets for your summer trip a few days ago. Did you manage to get that sorted?” as shown in Figure~\ref{fig: Scenario 2 - evening}. To achieve this, we will build an LLM prompting strategy that randomly selects a goal related to productivity and task management. 
Based on the literature and brainstorming, we developed a set of goals that the conversational agent can have per daily conversation; the goals are listed in Table~\ref{tab:goalsTable}. The morning check-in question and evening reflection prompt will be generated by creating a prompt with a randomly selected goal from the set.
The goal set will have two categories of goals, one focusing on productivity and the other focusing on mental well-being with regard to their work. We aim to design a system that can be personalized to each user and allow users to specify their preference between productivity and well-being so that the random selection will be weighted based on the users' selection.

We will also explore a crowd-sourcing approach to add new goals to the prompt model. Based on user interaction, a conversational agent will explicitly ask to specify their goal related to productivity and mental well-being. Then, the identified goal will be added to the goal table for the system to later use to generate inquiries. Added goals can be used for other users as well. Leveraging crowd-sourcing, we can increase the social value of the system and help users get a wider perspective of everyday life~\cite{HongAutismCrowdSource}.

\begin{table*}[t!]
    \centering
    \caption{Goals used in prompt design to generate questions from the LLM}
    \label{tab:goalsTable}
    \begin{tabular*}{\linewidth}{@{\extracolsep{\fill}}p{0.15\linewidth}p{0.15\linewidth}p{0.60\linewidth}@{}}
        \textbf{Prompt Focus} & \textbf{Theme} & \textbf{Goals} \\
        \midrule
         \multirow{5}{2cm}{Productivity}  &  Externalization  &  Help a worker externalize their plan on one task that they want to finish today. \\
         & Focus &   Help a worker find available time for them to focus without distraction.  \\
          &  Prioritization  &   Help a worker differentiate an urgent task from an important task in their daily plan.    \\ 
          &  Collaboration &  Help a worker understand the importance of collaboration and delegation among the daily tasks.    \\
          &  Reflection  &  Help a worker reflect on whether their daily plan matches their core values. \\
        \hline
        \multirow{6}{*}{Well-being}  &  Awareness  &  Help a worker be aware of their emotional state and understand how that can impact their day on a particular day.     \\
          &  Breaks  &   Help a worker plan their breaks for the day, which will help them manage their time efficiently.   \\
          &  Positivity &  Help a worker have a positive beginning of the day.     \\
          & Activity & Help a worker find time for an active lifestyle during the day at work.    \\
          & Stress management & Help a worker identify and regulate the stress that they may have for the day.  \\
        \hline
    \end{tabular*}
\end{table*}



\section{ONGOING EFFORTS}
In this section, we lay out our ongoing work in terms of the software development, user study design, and evaluation of the system. 

\subsection{Development}
Currently, we are developing a Slackbot~\cite{slackSlackYour} using SlackAPI~\cite{slackSlackYour} along with OpenAI GPT 3.5 model to incorporate the design goals described in the section above. 
We are developing the intial prototype in Slack ~\cite{slackSlackYour} because it is an application centered in the productivity domain. We are also exploring various prompting strategies to implement our design goals related to reflection and rotation. An initial strategy is to design the prompts from a goals-focused perspective where the LLM is provided with a random goal from Table~\ref{tab:goalsTable} and the conversation is centered around that goal.
\subsection{User study}
In order to answer our research questions and seek ecologically valid insights, we need to isolate and analyze the effect of rotation and personalized reflection as independent variables. In that vein, we plan to conduct a field deployment within-subjects study with two versions of ProductivityPartner:

\begin{itemize}
    \item The baseline version, which only supports static intervention.
    \item The full system, which has a rotation strategy implemented in the prompts along with the context-aware features.
\end{itemize}
During the study, participants will receive question prompts at the beginning and end of their working days over the course of some weeks delivered through Slack which can be accessed using a Desktop or Mobile phone. We also plan to distribute daily surveys based on the user condition which can include quantitative Likert-scale questions that measure participants' perceived productivity (“I was productive on the day when I used ProductivityPartnerBaseline/ProductivityPartnerFullSystem”) or perceived effectiveness of the system they used (“It was helpful to use ProductivityPartnerBaseline/ProductivityPartnerFullSystem to reflect and better plan my day”). On top of this, we can distribute weekly surveys with questions related to longitudinal reflection over the past week since it might take time for users to perceive changes in their productivity. After the field study, we aim to conduct semi-structured interviews to understand the concerns and perceptions users had while using the system.

\subsection{Evaluation}

In order to evaluate the results of our study, we will do a qualitative analysis of the dialogue exchanges~\cite{HuScreenTrack}. We will assess engagement by analyzing the logged participant interactions, such as the number of dialogues responded to, the time until a response was made, as well as the length, and content of responses. Moreover, we will perform quantitative analysis by performing repeated-measure ANOVA on the Likert-scale data gathered from the surveys.

\bibliographystyle{ACM-Reference-Format}
\bibliography{sample-base}


\end{document}